# MOBILE AGENTS FOR CONTENT-BASED WWW DISTRIBUTED IMAGE RETRIEVAL

Sabu .M Thampi[1], Dr. K. Chandra Sekaran[2]

[1] Assistant Professor, L.B.S College of Engineering, Kasaragod, Kerala, India
{smlbs@yahoo.co.in}

[2] Professor & Head, National Institute of Technology Karnataka, India
{kch@nitk.ac.in}

Abstract: At present, the de-facto standard for providing contents in the Internet is the World Wide Web. A technology, which is now emerging on the Web, is Content-Based Image Retrieval (CBIR). CBIR applies methods and algorithms from computer science to analyse and index images based on their visual content. Mobile agents push the flexibility of distributed systems to their limits since not only computations are dynamically distributed but also the code that performs them. The current commercial applet-based methodologies for accessing image database systems offer limited flexibility, scalability and robustness. In this paper the author proposes a new framework for content-based WWW distributed image retrieval based on Java-based mobile agents. The implementation of the framework shows that its performance is comparable to, and in some cases outperforms, the current approach.

## INTRODUCTION

During the last decade, we have seen a rapid increase in the size of digital image collections. As the computational power of both hardware and software have increased, the ability to store more complex data types in databases, such as video, audio and images has been drastically improved. These new media types offer other challenges, and demand a different treatment than pure text. Research in this area started in the 70's, based on traditional information retrieval, and is today both an active and important field in information and data management.

The World Wide Web is rapidly being accepted a universal access mechanism for network information. The popularity of Web suggests that web browsers may offer a compelling end-user interface for a large class of applications including Image Retrieval. A technology, which is now emerging on the Web, is Content Based Retrieval, CBR. A content-based query matches examples or prototypes to known instances of a certain media type based on a measure of similarity. For efficiency, similarity measures are frequently computed on sets of discriminant features (so called feature vectors) being extracted a-priori from stored media.

The present approaches towards supporting a market of digital images suffer from a number of disadvantages. All images have to be transferred across the network to some indexing process. Image gatherers request images using HTTP requests. Therefore, the gatherer needs to wait between requests, which increases the time required for requesting and receiving an image. This time is identified as a performance bottleneck. Hence an alternative architecture is proposed for distributed indexation and searching of images. This combines Content Based Image Retrieval technology, mobile agent technology and digital water marking. The widespread use of Java in network centric computing, attributed mainly to its global portability and security control system, gives Java the lead in Client/ Server programming and mobile computing.

The proposed framework, called the Image-Aglet framework utilizes the newest technology of mobile agents and CBIR and demonstrates its effectiveness over a specific application context (i.e. Image Retrieval). The main advantage of the framework is it frees the remote client to perform other more essential tasks. The implementation of the framework shows that the performance of the system is comparable to current approaches.

## MOBILE AGENTS

The term software agents refer to programs that perform certain tasks on behalf of the user. Software agents can be classified as static agents and mobile agents. Static agents achieve the goal by executing on a single machine. On the other hand, mobile agents migrate from one computer to another and executes on several machines. Mobility increases the functionality of the mobile agent.

A mobile agent consists of the program code and the program execution state. Initially a mobile agent resides on a computer called the home machine. The agent is then dispatched to execute on a remote computer called a mobile agent host. When a mobile agent is dispatched the entire code of the mobile agent and the execution state of the mobile agent is transferred to the host. The host provides a suitable execution environment for the mobile agent to execute. Another feature of mobile agent is that it can be cloned to execute on several hosts. Upon completion, the mobile agent delivers the results to the sending client or to another server.

Aglet Technology is a framework for programming mobile network agents in Java developed by the IBM Japan research group. The IBM's mobile agent is called 'Aglet', is a lightweight Java object. One of the main differences between an aglet and the simple mobile code of Java applets is the itinerary or travel plan that is carried along with the aglet. By having a travel plan, aglets are capable of roaming the Internet collecting information from many places.

An aglet can be dispatched to any remote host that supports the Java Virtual Machine. This requires from the remote host to pre-install Tahiti, a tiny aglet server program implemented in Java and provided by the Aglet Framework. To allow aglets to be fired from within applets, the IBM Aglet team provided the so-called "FijiApplet", an abstract applet class that is part of a Java package called "Fiji Kit". FijiApplet maintains some kind of an aglet context. From within this context, aglets can be created, dispatched from and retracted back to the FijiApplet.

# CONTENT-BASED IMAGE RETRIEVAL (CBIR)

Digital images and videos have an increasingly important role in today's telecommunication and our everyday life in information society. The past few years witnessed a proliferation of content-based image retrieval techniques. The key issue in CBIR is how to match two images according to computationally extracted features. Typically, the content of an image can be characterized by a variety of visual properties known as features. Most CBIR techniques fall into two categories: manual and computational. In manual approaches, a human expert may identify and annotate the essence of an image for storage and retrieval, computational approaches, on the other hand, typically rely on feature-extraction and pattern-recognition algorithms to match two images.

**CBIR using Gabor Texture features**

This section describes an image retrieval technique based on Gabor texture features. Texture is an important feature of natural images. Gabor filter (or Gabor wavelet) is widely adopted to extract texture features from the images for image retrieval and has been shown to be very efficient. Basically, Gabor filters are a group of wavelets, with each wavelet capturing energy at a specific frequency and a specific direction. Expanding a signal using this basis provides a localized frequency description, therefore capturing local features/energy of the signal. Texture features can then be extracted from this group of energy distributions. The scale (frequency) and orientation tunable property of Gabor filter makes it especially useful for texture analysis. A rotation normalization method that achieves rotation invariance by a circular shift of the feature elements so that all images have the same dominant direction is proposed here.

**Texture representation**

After applying Gabor filters on the image with different orientation at different scale, we obtain an array of magnitudes:

$$E(m, n) = \sum_x \sum_y | G_{mn}(x,y) |,$$
$$m = 0, 1 \ldots M\text{-}1; n = 0, 1 \ldots N\text{-}1$$

These magnitudes represent the energy content at different scale and orientation of the image.

The main purpose of texture-based retrieval is to find images or regions with similar texture. It is assumed that we are interested in images or regions that have homogenous texture, therefore the following mean $\mu_{mn}$ and standard deviation $\sigma_{mn}$ of the magnitude of the transformed coefficients are used to represent the homogenous texture feature of the region:

$$\mu_{m,n} = E(m,n) / P \times Q$$

$$\sigma_{mn} = \frac{\sqrt{\sum_x \sum_y (|G_{mn}(x,y)| - \mu_{m,n})^2}}{P \times Q}$$

A feature vector **f** (texture representation) is created using $\mu_{m,n}$ and $\sigma_{mn}$ as the feature components. Five scales and 6 orientations are used in common implementation and the feature vector is given by:

$$\mathbf{f} = (\mu_{00}, \sigma_{00}, \mu_{01}, \sigma_{01} \ldots \mu_{45}, \sigma_{45}).$$

**Rotation invariant similarity measurement**

The texture similarity measurement of a query image $Q$ and a target image $T$ in the database is defined by:

$$D(Q, T) = \sum_m \sum_n d_{mn}(Q,T)$$

Where

$$d_{mn} = \sqrt{(\mu^Q_{mn} - \mu^T_{mn})^2 + (\sigma^Q_{mn} - \sigma^T_{mn})^2}$$

Since this similarity measurement is not rotation invariant, similar texture images with different direction may be missed out from the retrieval or get a low rank.

A simple circular shift on the feature map is sufficient to solve the rotation variant problem associated with Gabor texture features. Specifically, we calculate total energy for each orientation. The orientation with the highest total energy is called the dominant orientation/direction. We then move the feature elements in the dominant direction to be the first elements in f. The other elements are circularly shifted accordingly. For example, if the original feature vector is *"abcdef"* and *"c"* is at the dominant direction, then the normalized feature vector will be *"cdefab"*. This normalization method is based on the assumption that to compare similarity between two images/textures they should be rotated so that their dominant directions are the same.

## THE IMAGE – AGLET FRAMEWORK

Specifically, the following components are needed to support Image-Aglet framework:
- An Image-applet

The Image-applet is responsible for forming a graphical client database interface that user can utilise to enter image requests. The Image-applet is an extension of the abstract FijiApplet class.
- An Image-Aglet

The Image-Aglet is created within the context of the Image-applet and is responsible for carrying the users request to the Broker server through the Web server, executing it, and returning the results back to the Image-applet context. The Image-Aglet is a Java-based extension of the Aglet class.

Other components associated with the system are image-index agent, image-search agent, parked agent and messenger agent.

### Demonstrating the Web-based Image-Aglet Framework

In this framework an extension of the client/server model, called the client/agent/server model is used. An agent is placed on the path from the client to the server. Any communication between the client and the server goes through the agent. This agent is a service-specific surrogate of the client, which is parked at the Broker Server, and it is maintained there for the duration of the application. Between the parked aglet and the remote client, another aglet carries requests and results back and forth.

A web-based Image-Aglet infrastructure has been set up to demonstrate the system (see Figure 1). The Image-applet and Image-Aglet are programmed and included in an HTML page at the web server machine. Aglet router is installed at the Web server machine. Additionally the Tahiti Aglet servers are installed in the Broker server as well as in the image Providers' servers.

To demonstrate first download at the Client host the html page containing the image-applet and the Image-Aglet. Through the image-applet's GUI a query is entered. Two Image-Aglets are fired from the Aglet-applet. The first one is called *parked Image-Aglet*.

The parked Image-Aglet carries the following message directions:
- The address of the URL where the broker server is located.
- Query to be executed at the broker server.
- The appropriate certificates for the aglet to be trusted at the broker server.

The role of the parked agent is to camp at the broker server's agent context, submit the client's request, load the appropriate drivers and collect the answer for the query. The second Image-Aglet is called the messenger aglet. The messenger aglet is responsible for carrying the result back to the Image-Aglet (see figure 1). Any subsequent requests are transmitted via the messenger aglet to the parked Image-Aglet.

As soon as the Broker server receives user request it sends image-index agents to the image providers' servers. Multiple copies of index agents may be created to visit servers of image providers. The index agents transport the texture feature extraction and collection algorithms. Index agents may compute and collect indexes of multiple images archives, which are sent back and merged to the broker's main index.

The client receives a list of image descriptors of images matching the query in the order of similarity. Each image descriptor consists at least of a thumbnail, the image identifier of that image (unique within the domain of the image provider), a measure of "similarity" to the original query image, and the URL of the Provider's agent server from which the image can be retrieved. The image themselves must be retrieved from the providers. This ensures that providers may identify customers and may apply digital watermarks to retrieved images. While retrieving thumbnails is free, retrieving full images is subject to access control. In accordance to agreements between brokers and providers, index agents are authorised to read full images for the purpose of indexing.

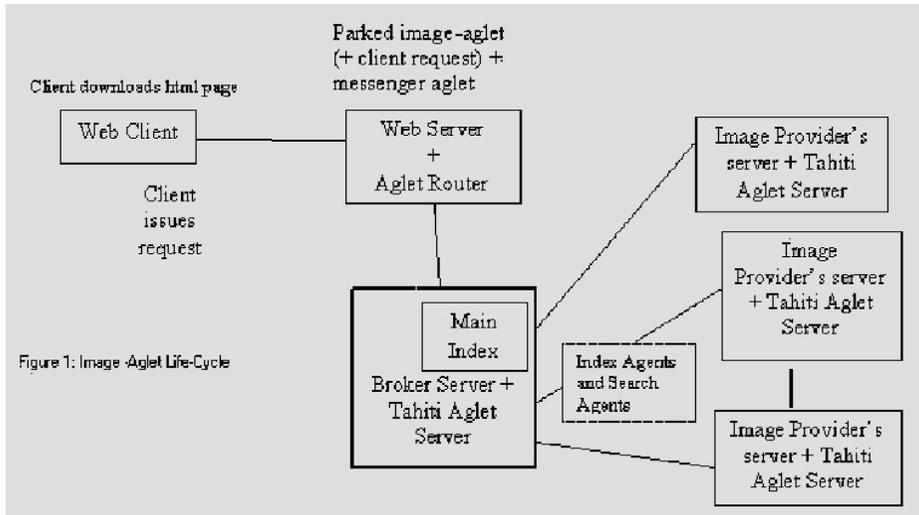

**Figure 1**: Image-Aglet Life-Cycle

After viewing the thumbnails of images, user can send requests to retrieve the images through the applet. The messenger aglet carries these requests to the Web server. The broker server creates a new type of mobile agents called search agents. Multiple copies of search agents are created to visit Providers' sites. Search agents are allowed to retrieve full images only if appropriate licenses were purchased beforehand. Digital watermarks are attached with the retrieved images. Search agents utilises the data kept in the main index of the Broker server.

**Refining Image-Aglet Framework**

The time required for the messenger aglet to travel between the image-applet and the parked Image-Aglet carrying results to the one way and new queries to the other create an overhead to the proposed system. Replacing the messenger aglet with two types of messages can eliminate this overhead. The first kind of message is delivered from the Image-Aglet to the image-applet contains the results of last query. The other message from the image-applet to Image-Aglet contains new client query. This method outperforms the traditional applet approach for retrieving images.

# PERFORMANCE EVALUATION

The performance evaluation compares the total time required by a web client to access and query image databases between the traditional applet-based and Image-Aglet methodologies. For each methodology the client is accessing the Web server via a 64 kbps dial-up connection to an ISP (BSNL). The tests were performed several times with more than 100 queries.

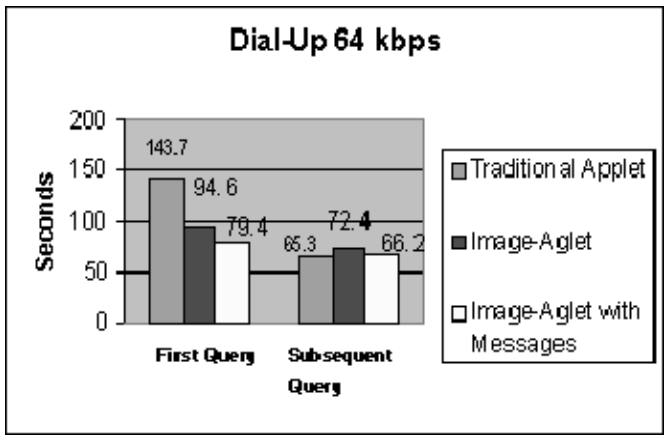

**Figure 2:** Mean times for 64 kbps client connectivity

The first step of data analysis was to perform a descriptive statistical analysis. This analysis gave information about the behaviour of the data sets involved in the statistical analysis, including a description of the mean, the median, and the standard deviation. In particular, a comparison of the means and standard deviations has been performed. This indicated that the most efficient methodology seems to be "parked Image-Aglet with messages" for the first query and the traditional applet method for subsequent queries. This is shown in Figure 2. However, the marginal difference between the performances of the three methods for subsequent queries is compensated by the significant difference in the performance for first queries, suggesting that the parked Image-Aglet with messages could be considered as the most efficient methodology for all cases of client connectivity.

# CONCLUSIONS AND FUTURE WORK

In this paper a new approach for developing client/server applications on the Web using Java mobile agents for Image retrieval is introduced. The mobile agent approach has a number of advantages. Image indexing is decentralized. It is computed "near" the image database by index agents migrating to the server of image providers. Images must not be transported across networks for index generation any more. Retrieved images can be personalized by watermarking them with identity of the purchaser.

The system can be further improved by extending the CBIR algorithm to extract more complex features of the images. This includes identification of objects and scenes from images.